\documentclass[12pt]{article}
\usepackage{a4wide}
\usepackage{amsmath}
\usepackage{amssymb}
\usepackage{graphicx}
\numberwithin{equation}{section}

\newcommand{\be}{\begin{equation}}
\newcommand{\ee}{\end{equation}}
\newcommand{\ba}{\begin{eqnarray}}
\newcommand{\ea}{\end{eqnarray}}
\newcommand{\no}{\nonumber}

\def\identity{1\!\!1}
\def\Re{\mathop{\mathrm{Re}}}

\def\Z{\mathbb{Z}}
\def\C{\mathbb{C}}
\def\R{\mathbb{R}}
\def\N{\mathcal{N}}
\def\M{\mathcal{M}}
\def\CP{\mathbf{P}}
\def\P{\mathbf{P}}
\def\vev#1{\langle#1\rangle}
\def\Sp{\mathop{\mathrm{Sp}}}
\def\identity{\mathrm{id}}
\def\Ric{\mathrm{Ric}}

\begin{document}

\begin{titlepage}
\begin{flushright}
UT-05-14\\
NSF-KITP-05-79\\
hep-th/0510061
\end{flushright}
\vfil
\begin{center}
{\LARGE\bf Distribution of Flux Vacua around\\[4mm]
Singular Points in Calabi-Yau Moduli Space}\\
\bigskip
\bigskip
\bigskip
{\Large Tohru Eguchi$^1$ and Yuji Tachikawa$^{1,2}$}\\

\vskip1cm 

\textit {$^1$\ {\large Department of Physics, Faculty of Science,}}\\
\textit {\large{University of Tokyo, Tokyo 113-0033,  JAPAN}}\\

\vskip1cm

\textit {$^2$\ \large{Kavli Institute for Theoretical Physics,}}\\
\textit {\large{University of California, Santa Barbara, CA 93106 , USA}}\\
\bigskip
\end{center}
\bigskip\bigskip\bigskip\bigskip\bigskip\bigskip
\centerline{\large\textbf{abstract}}
\bigskip
We study the distribution of type IIB flux vacua
in the moduli space
near various singular loci, e.g. conifolds, ADE singularities on ${\bf P}^1$, Argyres-Douglas point etc, using the Ashok-Douglas
density $\det (R+\omega)$.
We find that the vacuum density is integrable around each of them, irrespective of
the type of the singularities.
We study in detail an explicit example of an
Argyres-Douglas point embedded in a compact Calabi-Yau manifold.
\end{titlepage}

\section{Introduction}

Recently the problem of moduli stabilization and the landscape of flux vacua in string theory are receiving a great deal of attentions \cite{BussoPolchinski,GiddingsPolchinski,KKLT,statistics} (for an earlier reference, see \cite{Sethi}).
In the case of type IIB string theory, for instance,  a non-zero superpotential is generated and complex structure moduli  become fixed at its extremum
when one introduces RR and NSNS 3-form fluxes passing through cycles of the Calabi-Yau (CY) manifold. K\"ahler moduli may also be fixed when CY manifold obeys a certain topological condition
and D-brane instanton amplitudes
become non-vanishing\cite{KKLT,BetterRacetrack,BarrenLandscape}\footnote{
There are some developments to uncover the effect of the fluxes to
the generation of instanton amplitudes\cite{TripathyTrivedi,KalloshIndex,LustTripathy}. 
} . 
It turns out, however,  that such a mechanism of moduli stabilization leads to an  enormously large (possibly infinite) number of vacua in type IIB string theory. Problem of vacuum selection is now becoming one of the most challenging issues in superstring theory.

Actually the flux vacua are not distributed uniformly over the whole of moduli space of CY manifolds. Instead they are sharply peaked around the singular loci in the CY moduli space, such as the conifold points
\cite{AshokDouglas,DenefDouglas}\footnote{Mathematically-oriented readers will find quite helpful the discussions in \cite{Zelditch} and references therein.}.
It is well-known that various interesting non-perturbative phenomena take place at these singular points; i) appearance of massless matter at the conifold point\cite{BlackCondensation},
ii) generation of non-Abelian gauge symmetry at the ADE-singularities (fibered over ${\bf P}^1$) \cite{GeometricEngineering},
iii) emergence of mutually non-local massless solitons and scale invariance at Argyres-Douglas
point\cite{AD,NewSCFT,EHIY} etc.
Thus it seems worthwhile to study in some detail the behavior of flux vacua around the singular loci of CY moduli space.

In this paper we will use the density formula $\det(R+\omega)$ of Ashok and Douglas \cite{AshokDouglas}
($R$ and $\omega$ denotes the curvature and the K\"ahler form of the moduli space) 
for the flux vacua in the type IIB string theory and estimate the behavior of vacuum density
near Calabi-Yau singularities.
We shall show that the following behavior
\be
\mbox{vacuum density}\approx \frac{dz\,d\bar{z}}{ |z|^2(\log |z|)^{p}},
\qquad p:\mbox{positive integer, $p\ge 2$}
\ee
holds universally irrespective of the nature of singularities (conifolds, $\P^1$
fibration of ADE singularities, Argyres-Douglas point etc.) 
with  the value of integer $p$ depending on the specific cases.
Here $z$ denotes a coordinate in moduli space with the singularity located at $z=0$. The above formula
has been known in the case of conifold singularity,
and it is also verified numerically in some cases \cite{Taxonomy,ConlonQuevedo}.

We note that the vacuum density is in fact normalizable
\be
\int \mbox{vacuum density} < \infty,
\ee which roughly states that there exist only a finite number of vacua
concentrated around each singular locus.
This seems to be a further result suggesting the finiteness of the number of 
vacua in the type IIB string landscape (see a very recent discussion in \cite{DouglasLu}).

The structure of the paper is as follows: in section 2,
we review the Ashok-Douglas formula and necessary ingredients.
The discussion will be brief and is mainly to fix the convention.
In section 3, we lay out the basic strategies of studying the
growth of the Ashok-Douglas density near the singular locus in the
moduli space. 
The actual analysis is done in section 4.
In section 5, we take a specific
Calabi-Yau to examine a concrete example, and study 
the geometric-engineering limit and the Argyres-Douglas points.
We conclude in section 6 with some discussion.

Note Added: During the preparation of the manuscript,
we have noticed a very short announcement in
the mathematics archive \cite{LuNatsukawa} where results related to ours are discussed.

\section{Preliminaries}

In this section we  introduce our notations and conventions and
review the Ashok-Douglas formula.

\subsection{K\"ahler geometry}
We denote by $K$ the K\"ahler potential of the manifold.
In supergravity, the K\"ahler manifold  of the sigma model target space
needs to be a Hodge manifold, that is,
its K\"ahler form must be equal to
the curvature of some holomorphic line bundle $L$ 
in the Planck units\cite{PlanckQuantization}.
A holomorphic section $s$ of the bundle $L^{\otimes p}$
transforms as $s\to e^{pf}\,s$ 
under  the K\"ahler transformation $K\to K+f+\bar f$. 
The superpotential of a supergravity theory
is a holomorphic section of $L^{-1}$.
The covariant derivatives for the section of $L^{\otimes p}$ are given by
\ba
&&D_i = \partial _i -p (\partial_i K),\label{cov}\\
&&\bar D_{\bar i}=\bar\partial_{\bar i}.
\ea
Here $\partial_i=\partial/\partial z_i$ and $z_i$ denotes the coordinates
on the manifold. 
If one considers in general a Hermitean vector bundle $E$ with a connection $A=A_idz^i$
($A_i=-p\,\partial_i K$ in (\ref{cov})), 
the curvature form is defined as \begin{equation}
F=(dz^i D_i + d\bar z^{\bar i} \bar D_{\bar i})^2= 
\bar\partial A.
\end{equation}

The metric $g_{i\bar{j}}$ of a K\"ahler manifold is given by $\partial_i\partial_{\bar{j}}K$ and the K\" ahler form is defined by\begin{equation}
\omega= i dz^i \wedge d\bar z^{\bar j} g_{i\bar j} = i \partial \bar\partial K.
\end{equation}Then, the first Chern class of $L^{\otimes p}$
is given by \begin{equation}
c_1(L^{\otimes p})=  \frac {-ip}{2\pi } \bar \partial \partial K  = 
\frac{p \,\omega}{2\pi}.
\end{equation}

For the holomorphic tangent bundle, the connection one-form is  \begin{equation}
A=g^{i\bar k }\partial g_{j\bar k}
\end{equation}
Components of the Riemann tensor are given by 
\begin{equation}
R_{i\bar j k\bar l}=
 g_{n\bar l}\,\partial _{\bar j} g^{n\bar m} \partial_k g_{i\bar m}.
\end{equation}
Curvature two-form is then written as \begin{equation}
R=i g^{m\bar l} R_{i\bar j k\bar l} \,d\bar z^{\bar j}\wedge dz^k. 
\end{equation}

Thus the top Chern class of the bundle $L^{-1}\otimes \Omega^{(1,0)} {\cal M}$ is
\begin{equation}
 \det (  -\frac{ R}{2\pi}-\frac \omega{2\pi} ).
\end{equation}
When the bundle is positive, it calculates the number of zeros of its generic section.

\subsection{Special K\"ahler geometry}
Compactification of Type IIB string on Calabi-Yau with fluxes
inherits certain properties from compactification without flux.
Thus, part of the structure of the $\N=2$ supergravity in four dimension
comes into play.  Let us recall the structure of the vector multiplet scalars
in the $\N=2$ supergravity. We denote the number of the vector multiplets
by $n$.

Let $n$  complex scalars $z_i$, $i=1,\ldots, n$
parametrize the scalar manifold $\M$.
Fundamental data defining the special K\"ahler geometry
is the flat vector bundle $V$ of rank $2n+2$
equipped with the symplectic pairing $\eta^{IJ},\, I,J=1,2,\ldots,2n+2$ 
and the holomorphic section
$\Omega=\{\Omega_I\}$
of the bundle $L^{-1}\otimes V$.
$\Omega_I$ are called as the periods, or the (projective) special coordinates
of the manifolds.  We sometimes abbreviate the pairing
$\eta^{IJ}A_I B_J$ of two vectors as $\eta AB$ if no confusion would arise.

The periods are constrained by the  transversality condition
\begin{equation}
\eta^{IJ}\Omega_I \partial_i \Omega_J=0.\label{transv}
\end{equation}
One can recover the prepotential $F$ in the following way:
Introduce another variable $z_0$ parametrizing $\C$ and consider a manifold $\C\times \M$.
Define functions $\Omega_I(z_0,z)$ on $\C\times \M$  by $\Omega_I(z_0,z)=z_0 \Omega_I(z)$.
Let us note that eq. \eqref{transv} holds also on $\C\times \M$, including $\partial_i=\partial_{z_0}$.
We take a canonical symplectic basis and split $\Omega_I$ into $(X^\Lambda,F_\Lambda)$
($\Lambda$ runs from $0$ to $n$).
We introduce a function $F$ by the formula $F=X^\Lambda F_\Lambda/2$.
$\C\times\M$ can be parametrized by $X^0,\ldots,X^N$. Then
\eqref{transv}  in this coordinate system reads \begin{equation}
F_\Sigma \partial_\Lambda X^\Sigma - X^\Sigma\partial_\Lambda F_\Sigma=0,
\qquad\text{i.e.}\quad  F_\Lambda=X^\Sigma\partial_\Lambda F_\Sigma.
\end{equation}
Thus we have\begin{equation}
\frac{\partial F}{\partial X^\Lambda}=\frac{F_\Lambda}2+
\frac12 X^\Sigma \partial_\Lambda F_\Sigma=F_\Lambda.
\end{equation}

Standard formulae of special geometry\cite{StromingerSpecial,CernSpecial}
are then given by
\begin{align}
K&=-\log i\eta^{IJ} \bar \Omega_I \Omega_J,\\
e^{K}&=1/(i\eta\bar\Omega\Omega),\\
g_{i\bar j}&=-\frac{\eta\partial_{\bar j}\bar \Omega \partial_{i} \Omega}{(\eta \bar \Omega\Omega)}
+\frac{(\eta\bar\Omega\partial_i\Omega)(\eta\bar\partial_{\bar j}\bar\Omega\Omega)}
{(\eta \bar \Omega\Omega)^2},\\
F_{ijk}&=\eta^{IJ} \Omega_I  \partial_i \partial_j \partial_k \Omega_J
=-\eta^{IJ}\partial_i\Omega_I\partial_j\partial_k\Omega_J,\\
R_{i\bar jk\bar l}&=g_{i\bar j}g_{k\bar l}+g_{i\bar l}g_{k\bar j}-e^{2K} g^{m\bar n}F_{ikm}
\bar F_{\bar j\bar l\bar n}.
\end{align}
Note that the $\Omega_I$ may be multivalued because of the
 holonomy or monodromy  of $V$ in $\Sp(2n+2,\Z)$.
$F_{ijk}$ and $R_{i\bar{j}k\bar{l}}$ are invariant under the monodromy transformation.

\subsection{Flux superpotential and Landscape}\label{fluxlandscape}
The space of complex structure moduli $\M$ of a CY threefold $X$
is a special K\"ahler manifold of complex dimension $n=h^{1,2}(X)$.
The periods $\Omega_I$ are given by
\begin{equation}
\Omega_I=\int_{C_I}\Omega(z_i)
\end{equation}where $C_I$ is a basis of three-cycles in $X$
and $\Omega$ is the holomorphic three-form of $X$.
$\Omega$ depends on the moduli of the Calabi-Yau manifold.
The fact $\Omega$ is defined only up to the scalar multiple
depending on the moduli
is reflected to the fact that the K\"ahler  transformation acts on $\Omega_I$.

Let us compactify type IIB theory on $X$.
When one introduces NSNS and RR fluxes $H^{NSNS},H^{RR}$,  one obtains a four-dimensional 
$\N=1$ supergravity theory with a superpotential
\begin{equation}
W=\eta^{IJ}(\tau H^{NSNS}_I +H^{RR}_I)\Omega_J,\label{GVW}
\end{equation}where $\tau$  is the axiodilaton
and $H^{NSNS}_I,H^{RR}_I$ are the integral fluxes passing through 
the three-cycles $C_I$ \cite{GVW,TaylorVafa}.

The flux superpotential \eqref{GVW}  generically
gives masses to the complex structure moduli and the axiodilaton.
One needs to introduce a certain number of D3 branes and orientifolding
in order to satisfy the tadpole cancellation condition, which in general puts restrictions on the 
allowed values of the fluxes. 

There are also the moduli for the K\"ahler deformation
in CY manifolds 
and for the complete moduli stabilization we also have to consider mechanisms to give masses to the K\"ahler moduli.
In this paper we restrict our attention to the sector of complex moduli of CY manifolds.

A number of choices is possible for the set of
integer fluxes $H^{NSNS}_I,H^{RR}_I,\,I=1,\ldots,2n+2$ and this is the origin of  the existence of enormous
number of flux vacua in type IIB theory.  We may study the statistical distribution of these vacua
in the moduli space\cite{statistics,AshokDouglas},
instead of studying them one by one.
We do not know the dynamical mechanism behind the
\textit{a priori} probability distribution of the fluxes $H_I^{NSNS},H_I^{RR}$.
A zero-th order approximation may be to assume that the fluxes
obey a Gaussian distribution. Then,  the correlators
among the superpotential at various points on the moduli are
given by \begin{align}
\vev{W(z_i)W^*(w_i)}&\propto e^{-K(z_i,w_i^*)},\label{GaussianCorrelation1}\\
\vev{W(z_i)W(w_i)}&=0=\vev{W^*(z_i)W^*(w_i)}.\label{GaussianCorrelation2}
\end{align}
From the point of view of the study of the distribution of vacua
under random superpotentials, 
we may concentrate on what can be derived from the basic correlation functions
\eqref{GaussianCorrelation1} and \eqref{GaussianCorrelation2}.

Supersymmetric AdS vacua are at the extrema of the superpotential,
$D_iW=0$.
The distribution function of such vacua
over the moduli space  is given by
\begin{equation}
\rho(z,\bar z)=
\vev{\delta(D_i W(z))\delta(\bar D_{\bar \imath} W(\bar z)^*)
\left|
\det \begin{pmatrix}
\partial_i D_j W& \partial_i D_{\bar \jmath}W^*\\
\partial_{\bar \imath} D_j W&\partial_{\bar \imath} D_{\bar j} W^*
\end{pmatrix}\right|}.
\end{equation} 
The absolute value is required to count each vacua with a weight one.
It is somewhat difficult to evaluate this quantity because of the
absolute value sign.  We may consider instead 
\begin{equation}
\tilde\rho(z,\bar z)=
\vev{\delta(D_i W(z))\delta(\bar D_{\bar \imath} W(\bar z)^*)
\det \begin{pmatrix}
\partial_i D_j W& \partial_i D_{\bar \jmath}W^*\\
\partial_{\bar \imath} D_j W&\partial_{\bar \imath} D_{\bar j} W^*
\end{pmatrix}}
\end{equation}which counts vacua with a sign.
Ashok and Douglas showed in \cite{AshokDouglas} that 
it is given by\begin{equation}
\tilde\rho(z,\bar z)\prod_idz^i\wedge d\bar z^{\bar\imath}
\propto \det(-\frac{ R^i{}_{j}}{2\pi}-\frac{\delta^i{}_{j}\omega}{2\pi}).
\label{AshokDouglasFormula}
\end{equation}
The right hand side is the determinant of the
curvature tensor of $\Omega^{(1,0)}{\cal M} \otimes L^{-1}$, of which $D_iW$ is the section.
Since $\tilde\rho$ gives the lower bound for the number of vacua $\rho$,
the formula tells that the vacua do not distribute uniformly. Rather, they tend to 
concentrate near the points where the curvature of the moduli space is peaked.
Thus, it is of importance to study the convergence property of
\eqref{AshokDouglasFormula} near various kind of
singularities in the moduli space, which is the main objective of our study.

\section{Strategy of the analysis}
In the following, we study the behavior of the Ashok-Douglas
density \eqref{AshokDouglasFormula} on the complex structure moduli
space alone, neglecting the contribution from the axiodilaton and other
scalar fields.  This is not going to be the full answer to our problem,
but is hopefully an important  step in that direction.

Before proceeding, we would like to mention
the relation of our work
with those in the mathematical literature.
When phrased in purely mathematical terms, our work 
studies the convergence properties of various products
of Chern classes and K\"ahler forms
near the singular loci of the CY moduli space.
If one chooses the $n$-th power of the K\"ahler forms as the integrand,
the problem becomes precisely the convergence of the volume
of the moduli space of Calabi-Yau
studied in \cite{Todorov,LuSun1}.
The integral of the products of the first Chern class and the K\"ahler form
was studied in \cite{LuSun2} and was shown to be finite.
What we need to study is basically the sum of all possible
Chern classes.
In \cite{Hayakawa,Wang}, the distance to the singular locus
was studied using a similar method.

\subsection{Singularities in Calabi-Yau moduli space}
As briefly reviewed in section \ref{fluxlandscape},
the moduli space $\M$ for the complex structure deformation of Calabi-Yau
three-fold $X$ is a special K\"ahler manifold. 
It is not smooth and compact, however.
It contains a number of so-called
discriminant loci, around which the periods $\Omega_I$
have non-trivial monodromies preserving the symplectic pairing $\eta$.
Each point on the complement of the discriminant loci
corresponds to the non-singular Calabi-Yau.
An important fact is that a deep mathematical theorem by
Viehweg\cite{Viehweg}
ensures that the moduli space $\M$ of smooth Calabi-Yau manifolds
$X$ is quasi-projective,
that is, it can be realized as $\overline {\M}\setminus D$
where $\overline{\M}$ is a subvariety of some projective space
and $D$ is a divisor.
Although some kind of singular Calabi-Yau manifolds should
correspond to the point of $D$,
it is not known in complete generality which kind of singular 
Calabi-Yaus to allow
in order to compactify $\M$ to $\overline{\M}$.

Components of $D$ intersect among themselves,
and they in general develop singularities on them.
For example, it is known that the Argyres-Douglas
point corresponds to a cusp of the discriminant locus,
as we will review below.

What greatly facilitates the analysis is that,
using Hironaka's theorem,
we can resolve the singularity of $D$
by a number of blow-ups 
so that the singular loci $D$
can be made to have `normal crossings', i.e. they intersect transversally. 
Thus, we can always take, near the intersection of
$k$ singular loci, 
a coordinate patch of the form
$(z_1,\ldots,z_n)\in(\Delta^*)^k\times \Delta^{n-k}$
where $\Delta$ is a unit disk and $\Delta^*=\Delta\setminus0$
is a punctured disk so that the loci themselves are given by
$z_1z_2\cdots z_k=0$. 

In understanding the divergence or the convergence of the curvature and 
the volume form near the discriminant loci, what we actually examine is the outside
of the singular locus; thus the construction above
can be seen as providing
a good coordinate system to study the singular loci.
For concrete examples of discriminant loci, we can indeed show
in each case that the loci can be made to have normal crossings by an appropriate 
blow up procedure.

\subsection{Monodromy and Schmid's nilpotent orbit theorem}
We are going to study the curvature of the moduli space.
As a special K\"ahler manifold, all the properties are 
encoded in the behavior of the periods $\Omega_I$.
Thus we first need a good control over the behaviors of $\Omega_I$
near the discriminant locus.
As the periods are holomorphic, their properties are basically
determined by the monodromies around the
discriminant locus.  We can study case by case how each 
monodromy type is related to a particular behavior of the curvature of moduli space. 
In the mathematical literature, however, a great deal is known on the generic properties
of the monodromy matrix and the behavior of the periods.

The proper mathematical
language  to use is the pure and mixed Hodge structures.
Although they are somewhat unwieldy, 
results from these subjects are essential.
We have collected  necessary definitions and the facts
in the appendix \ref{HodgeStructure} to fill the gap
between language used in  mathematics and in string theory.
We present below the necessary results we use
in the main part of the paper.

First, let us recall some terminologies.
A  matrix $N$ is called nilpotent  when there is some integer $k$
such that $N^k=0$.  A matrix $M$ is called unipotent when
$M-1$ is nilpotent.
A matrix $M$ is called quasi-unipotent if there is some integer $k$
such that $M^k$ is unipotent.

Traversing the circle around a singular locus
leads to some automorphism $\gamma$ of the $H^3(X,\Z)$
which preserves the intersection form. We call $\gamma$
the monodromy matrix. It is known that $\gamma$ is
quasi-unipotent. 
Thus there exists some integer $k$ and some nilpotent matrix $N$
such that $\gamma^k=\exp(N)$. $N$ is known to satisfy $N^4=0$.
The sketch of the proof is provided in the appendix \ref{monodromytheorem}.

Let us introduce a coordinate $z$ such that $z=0$ is the singular locus.
The nilpotent orbit theorem of Schmid says that (precise statement
is reproduced in the appendix \ref{VHS})
\begin{equation}
\Omega'\equiv \exp(-\frac N{2\pi i k}\log z)
\Omega, \hskip4mm (\Omega \mbox{   is a column vector made of periods } \Omega_I)
\end{equation}is analytic in $z^{1/k}$ in a suitable K\"ahler gauge
and starts with a nonzero constant term.
That is, there appear no negative powers of $z$ in the periods $\Omega'$.
Taking the coordinate change $z^{1/k}\to z$,
we can assume that there appear no fractional powers without
losing generality.

Let us suppose that a singular locus lies at $z=0$,
and expand the periods using the nilpotent orbit theorem as
\begin{equation}
\Omega_I=\sum_{i,j\ge 0} c_{Iij}z^i (\log z)^j.
\end{equation}
We will see that whether nonzero $c_{I,0,j}$ with $j>0$
exists or not affects
the behavior of various quantities
via the divergence or convergence of $e^{-K}$ at $z=0$.
Hence we will split the following discussions to two cases.
Case I is when all $c_{I,0,j}$ is zero for $j>0$;
Case II is when some of $c_{I,0,j}$ is nonzero $j>0$.
As we will see in the following,
conifold singularities and Argyres-Douglas singularities
are included in Case I , while the large complex structure 
and the geometric engineering limit are included in Case II.
Case I is treated in section \ref{CaseI} and Case II is treated in section \ref{CaseII}.

\section{Generic Estimate of the Growth}

After these preparations,
we study the growth of the Ashok-Douglas density $\det(R+\omega)$
near the singularity. 
First, note that \begin{equation}
\text{index density}\approx dz d\bar z \sqrt g 
\epsilon^{\overbrace{.......}^{2n}} \epsilon^{\overbrace{........}^{2n}} 
\overbrace{Q_{....}Q_{.....}\cdots  Q_{.....}}^{n}\label{QQ}
\end{equation}where $Q_{....}$ is a linear combination of $e^{2K} F_{...}\bar F_{...} g^{..}$ and $g_{..}g_{..}$. The dots denote various indices to be contracted.

Next, we rewrite $\epsilon^{....}\epsilon^{.....}$ by a totally-antisymmetrized 
product of $2n$ inverse metrics $g^{[.[.}g^{..}g^{..}\cdots g^{.].]}$.
Furthermore, we express $Q$ as a contraction of $g^{..}$ with a
linear combination of $ e^{2K}F_{...}\bar F_{...}$  and $g_{..}g_{..}g_{..}$. Then the density is written  schematically as\begin{equation}
\propto dz d\bar z\sqrt g \underbrace{\overbrace{g^{..}g^{..}\cdots g^{..}}^{3n}
\overbrace{(e^{2K}F_{...}\bar F_{...} +g_{..}g_{..}g_{..})\cdots(e^{2K}F_{...}\bar F_{...} +g_{..}g_{..}g_{..})
}^{n}}_{\text{denote this part by $I$}}\label{integrand}
\end{equation}
We expand the factor $I$ and rewrite it as a sum of terms of 
the form $g^{..}\cdots g^{..}e^{2mK}F_{...}\bar F_{...}\\
\cdots F_{...}\bar F_{...}g_{..}\cdots g_{..}$.
By contracting $g_{..}$ with $g^{..}$,
each term in $I$ is reduced to the form
\begin{equation}
g^{..}\cdots g^{..}e^{2mK}F_{...}\bar F_{...}\cdots F_{...}\bar F_{...}\label{terms}
\end{equation} without $g_{..}$ factors.
Here $m$ is the number of times the factors $F\bar F$ are taken in the expansion.
Thus, we need to estimate the growth of $e^K$, $F_{...}$ and $g^{..}$
and then to plug the result into \eqref{terms}.

\subsection{Conifold}\label{CaseI}
The symplectic pairing of $2n$ periods $\eta_{2n}$ splits into $\eta_2\oplus \eta_{2n-2}$,
and we assume a monodromy acting on the periods $\Omega_1,\Omega_2,\ldots,$ to be of the form \begin{equation}
\begin{pmatrix}
1&m\\
0&1
\end{pmatrix} \oplus \identity_{2n-2}
\end{equation}  This is the familiar case of the conifold singularity. $m$ is given by the number and the charges
of the multiplets which become massless at the singularity.

We take one of the special coordinates $a\equiv \Omega _2$ as a local coordinate and assume
that the locus is given by $a=0$.
Then, from the monodromy we deduce \begin{align}
\Omega_1&\approx \frac m{2\pi i} a\log a + \cdots,\\
\Omega_2&\equiv a,\\
\Omega_3&\approx \cdots,\\
\vdots&
\end{align}
where $\cdots$ denote analytic functions of $a$. There will be dependence on other coordinates,
but it must all be analytic. We will call this kind of behavior of the periods
as the conifold-type. This is case I in the classification given in the last section.
An important point is that the Yukawa coupling $F_{ijk}$ is monodromy invariant, thus it should contain no $\log a$ factors.

To facilitate the calculation we switch to a variable $t$ defined by $a=e^{-t}$ and the problem is to find the
divergence of various quantities near $\Re t\approx +\infty$. 
We denote the real and the imaginary part of $t$ by $r$ and $\theta$, respectively.
$t$ and $t+2\pi i$ should be identified.
Then the period behaves as
\begin{align}
\Omega_1&\approx - \frac m{2\pi i}  t e^{-t} + \cdots,\\
\Omega_2&\equiv e^{-t},\\
\Omega_3&\approx \cdots,\\
\vdots&
\end{align} Now the dots signify Taylor series in $e^{-t}$.

A fundamental result
that the mixed Hodge structure on the singularity is polarized
guarantees that $\eta\bar\Omega\Omega$ approaches a nonzero constant
at $r\equiv \Re t\to\infty$. 
This means that there are some periods $\Omega_3, \Omega_4, \ldots$
which remains nonzero in the limit $a\to 0$.
For further details, see appendix \ref{VHS}.

Let us estimate various quantities around the conifold locus.
We introduce letters $u,v,...$ to 
represent the coordinates other than $t$.
Indices $i,j...$ label, as before, the generic coordinates of the moduli space.

\paragraph{Estimate of $g_{i\bar j}$}
\begin{itemize}
\item 
Using the expansion above and the fact that $K$ is monodromy invariant,
$K$ has expansions like \begin{equation}
\sum_{k=0}^1\sum_{l,m\ge 0} c_{k,l,m}(t+\bar t)^k e^{- l t}e^{-m\bar t}.
\end{equation}
From the assumption, $c_{k,0,0}=0$ for $k>0$. On the other hand
it is known that $c_{0,0,0}\ne 0$ from the theory
of variation of Hodge structure, see appendix \ref{VHS}.
Thus $e^K$ has a nonzero limit when $r\to \infty$.
\item Hence we find \begin{equation}
g_{t\bar t}\sim r^p e^{-2r},\qquad
g_{t\bar v}\sim e^{-r},\qquad
g_{u\bar v}\sim 1
\label{met}\end{equation}
with $p=1$. We have introduced an integer $p$ 
so that it facilitates the generalization later.
\item Thus $dt d\bar t \sqrt{g} \sim r^{p}e^{-2r} dr d\theta$, which readily converges at $r\approx \infty$.
\end{itemize}

\paragraph{Estimate of $g^{i\bar j}$}
\begin{itemize}
\item Break $g_{i\bar j}$  into two parts: \begin{equation}
g_{ij}=g_{ij}^{(0)}+\delta g_{ij}
\end{equation}
where $g^{(0)}$ contains $g_{t\bar t}$ and $g_{u\bar v}$,
$\delta g$ contains $g_{t\bar v}$.
\item Using the formula 
\begin{equation}g^{-1}=g_{(0)}^{-1}-g_{(0)}^{-1} \delta g g_{(0)}^{-1}+\cdots,\end{equation}
we find  \begin{equation}
g^{t\bar t}\sim r^{-p} e^{2r},\qquad
g^{t\bar v}\sim r^{-p} e^{r},\qquad
g^{u\bar v}\sim 1.
\label{invmet}\end{equation} 
\end{itemize}

\paragraph{Estimate of $F_{ijk}$}
\begin{itemize}
\item $\partial_t \Omega_I$ at most behaves as $t e^{-t}$ and is exponentially small.
\item $\partial_u \Omega_I$ will be at most constant.
\item Second-derivatives of $\Omega_I$ 
are exponentially small
when they contain derivatives in $t$ and at most constant when do not.
\item $F_{ijk}=\eta^{IJ} \partial_i \Omega_I \partial_j\partial_k \Omega_J$.
Hence $F\sim e^{-t}$ if only one of $i,j,k$ is $t$, $F\sim e^{-2t}$ if two or three indices are $t$,
and $F\sim 1$ otherwise. There is no prefactor of $t$ because $F$ is monodromy invariant.
\end{itemize}

\paragraph{Estimate of index density}
\begin{itemize}
\item 
Recall the expansion \eqref{terms} and let us plug in the estimate
obtained so far.
Let us notice that to
each subscript $t$ or $\bar t$ there is a corresponding factor $e^{-r}$ (\ref{met}), 
and to each  superscript $t$ or $\bar t$  a factor $e^{r}$ (\ref{invmet}),
\textit{except for} $F_{ttt}$ and
its conjugate (they behave as  $e^{-2r}$).  Then, the summand in $I$ carries one positive power of exponential $e^{r}$
if it contains one $F_{ttt}$ (or its conjugate), and two positive powers of exponential $e^{2r}$
if it contains both $F_{ttt}$ and $\bar F_{\bar t\bar t\bar t}$.
Hence the leading contribution to $dtd\bar t \sqrt g I $ are the terms
containing $F_{ttt}$  and $\bar F_{\bar t\bar t\bar t}$. If we
recall the behavior of the volume factor  $dtd\bar t \sqrt g \approx drd\theta r^{p}e^{-2r}$, we find that the  exponential factors altogether cancel in the index density.
\item Finally let us show that the terms identified above 
 contain a sufficiently large number of negative powers of $r$.
 The structure of the term is \begin{equation}
 dtd\bar t \sqrt{g} F_{ttt}\bar F_{\bar t\bar t\bar t} 
 F_{...}\bar F_{...} \cdots F_{...}\bar F_{...} g^{..}\cdots g^{..}.
 \label{density2}\end{equation}
It turns out that the $F_{...}\,(\bar F_{...})$ appearing in (\ref{density2}) other than $F_{ttt}$ and 
$\bar F_{\bar t\bar t\bar t}$ carries at most one $t\,(\bar t)$ in the subscript,
because of the two epsilon symbols in \eqref{QQ}
 
The more we use $g^{t\bar t}$ in contracting indices,
the less suppression factors of $r^{-p}$ we have (\ref{invmet}).
Here there exist at least  six $t$ and $\bar{t}$'s from $F_{ttt}\bar F_{\bar t\bar t\bar t}$,
which  guarantees the
existence of at least a factor of $r^{-3p}$. 
By multiplying with $dt d\bar t \sqrt{g}$, we obtain
 \begin{equation}
\text{integrand} \approx dr d\theta r^{-2p}
 \end{equation} which converges at $r= \infty$. Q.E.D.
  
\end{itemize}

An easy generalization is possible
when the periods contains the terms of the form
\begin{equation}
\sim a^k(\log a)^m = e^{-kt} t^m
\end{equation} but
there are no `bare' $t=-\log a$ factors without being accompanied by powers of $a=e^{-t}$.
We call these type of degeneration as the `generalized conifold-type'.

$e^K$ can be expanded again in the form\begin{equation}
\sum_{k,l,m\ge 0} c_{k,l,m}(t+\bar t)^k e^{- l t}e^{-m\bar t}.
\end{equation}
Another condition is that $e^K$ is
the sum of the products of periods  and their conjugates,
which does not contain bare factors of $t$.
Thus when $c_{klm}$ is nonzero for $k>0$, $l$ and $m$ must be strictly positive.
There is a maximum for $k$ for which $c_{k,l,m}$ is nonzero;
we denote it by $p$. 
Then \begin{equation}
g_{t\bar t}\sim  r^{p} e^{-2r},\qquad
g_{t\bar v}\lesssim  e^{-r}
\end{equation} 
and the rest of the analysis above goes through unmodified.
We can also show that the integral of vacuum density converges when two or more
discriminant loci of generalized-conifold type intersect.




\subsection{Large complex structure limit}\label{CaseII}

Next we deal with the case where the periods have a bare $t=-\log a$ factor unaccompanied by
$e^{-t}$ 
around the discriminant loci.
Let us expand again the periods using the nilpotent orbit theorem,
\begin{equation}
\Omega_I=\exp(t \frac{N}{2\pi i})
(\Omega_I^{(0)}+\Omega_I^{(1)} a+ \Omega_I^{(2)}a^2+\cdots).
\end{equation} Let $p$ be the maximum integer such that
$N^p \Omega_I^{(0)}\ne 0$ for some $I$.
$i\eta\bar \Omega\Omega$ starts with the expansion
\begin{equation}
i\eta\bar \Omega\Omega =\sum_k c_k (t+\bar t)^k  + \text{exponentially small terms in $t$...} 
\end{equation} 
A basic result of the variation of Hodge structure is that 
$c_k$ is zero for $k>p$ and nonzero for $k=p$ (see appendix \ref{VHS}).
It is also known that $p\le 3$. 
Then $K \sim p\log (t+\bar t)$.
Thus \begin{equation}
g_{t\bar t}\sim 1/r^2,\qquad
g_{t\bar v}\sim 1/r
\end{equation}These estimates guarantee the convergence of the volume.
The inverse metric is \begin{equation}
g^{t\bar t}\sim r^2,\qquad
g^{t\bar v}\sim r.
\end{equation}

The rest of the discussion need to be done separately for
 $p= 3$, $p=2$ and $p=1$.

\begin{itemize}

\item $p=1$.
Then $F_{ttt}=\eta^{IJ}\Omega_I (\partial_t)^3\Omega_J$
and $F_{ttu}=\eta^{IJ}\Omega_I (\partial_t)^2\partial_u\Omega_J$
are exponentially small,
 because the derivatives with respect to $t$ kill the $O(t)$ and $O(1)$ factors.
Thus, the term which is not exponentially small
contains only $F_{tuv}$ type terms.
Hence the number of the subscript $t$ to be contracted
is at most equal to the number of $F$.
As each $F$ is accompanied by $e^K\sim 1/r$,
this cancels the positive factors of $r$ from the inverse metric.
Thus $I$ is bounded from above by a constant.

\item $p=2$.
Just as in the previous case, 
one finds $F_{ttt}$ to be exponentially small.
Thus, in the terms which are not exponentially small,
the number of the subscript $t$ to be contracted
by $g^{t\bar t}$ or $g^{t\bar v}$ 
is at most equal to twice the number of $F$.
This means the term is convergent,
 since each $F$ is accompanied by a factor $e^K\sim 1/r^2$.

\item $p=3$.
$F_{ijk}$ is at most constant as they are monodromy invariant 
and each is accompanied by a factor of 
$e^K
\sim  1/r^3$. 
Thus the convergence 
of the index density is  guaranteed.

\end{itemize}

\subsection{Landau-Ginzburg point}\label{CaseIII}
Lastly, we would like to discuss the behavior of the vacuum index density
around the Landau-Ginzburg (LG) point for completeness.
In the case of quintic hypersurfaces,
the monodromy $M$ satisfies $M^5=1$ at the LG point,
that is , the monodromy is idempotent.
Here we call any singular loci with idempotent monodromy
LG-type loci.

Suppose we have a singular locus with monodromy matrix $M$
which satisfies $M^p=1$.
Let the singular locus lie at $a=0$ with $a$ as the local coordinate
of the disk.
Changing variables to $z$ with the relation $a=z^p$, 
we obtain a variation of Calabi-Yau manifolds with trivial monodromy.
From the nilpotent orbit theorem,
the periods are analytic in $z$ and have the expansion \begin{equation}
\Omega_I=\Omega_I^{(0)}+\Omega_I^{(1)}z+\Omega_I^{(2)}z^2+\cdots.
\end{equation}with nonzero $\Omega_I^{(0)}$
and nonzero $i\eta \overline{\Omega_{(0)}}{\Omega_{(0)}}$.

If $\Omega_I^{(1)}\ne 0$, then nothing strange happens.
$g_{z\bar z}$ starts with nonzero constant, and $F_{...}$ is bounded.
Since the coordinate $z$ is just a $p$-fold cover of the coordinate $a$,
it guarantees the convergence of the volume, the vacuum index density and so on.

Although we do not know general arguments to show $\Omega_I^{(1)}\ne 0$,
we can check that it is indeed satisfied in many cases.
We suspect that it is a generic feature of LG points.

\section{Concrete Example}

Next we turn to the discussion of ADE type singularities fibered over ${\bf P}^1$,
which corresponds to the so-called geometric engineering of ${\cal N}=2$ SUSY gauge theory,
and then in particular the case of Argyres-Douglas point which can be reached by further fine-tuning of the parameters
in pure $SU(3)$ gauge theory.

Let us first describe briefly the structure of the Calabi-Yau manifold
we use as an example.
We take a three-modulus Calabi-Yau $X$, 
which is a type IIB dual of the so-called $STU$ model of heterotic compactification.
The following summary is based on ref. \cite{Billo}.
$X$ is described as
a degree $24$ hypersurface in the weighted projective space
$WCP_{24}(1,1,2,8,12)$,
which is invariant under the action of $\Z_{24}\times \Z_{24}$.
Denoting the homogeneous coordinates of the weighted projective space by
$[x_1:x_2:x_3:x_4:x_5]$, the identification under the group action is given by
\begin{equation}
[x_1:x_2:x_3:x_4:x_5]\sim
[stx_1:s^{-1}tx_2:t^{-2}x_3:x_4:x_5]
\end{equation}where $s,t$ are some 24th roots of unity.
The defining equation of the hypersurface is given by
\ba
&&0=\frac{B}{24} (x_1^{24}+x_2^{24})-\frac{1}{12}(x_1x_2)^{12}
+\frac1{12}x_3^{12}+
\frac13x_4^3+\frac12 x_5^2
-\psi_0 x_1x_2x_3x_4x_5\no \\
&&-\frac16\psi_1(x_1x_2x_3)^6-\frac14\psi_3(x_1x_2x_3x_4)^2-\frac14 \psi_4 (x_1x_2x_3)^4 x_4-\frac13\psi_5(x_1x_2x_3)^3x_5.
\label{defining}\ea
It is known that this space has the structure of a $K3$ fibration and $K3$ fiber itself has an elliptic fibration.
Thus it is relatively straightforward to construct  cycles
and compute their periods in this CY manifold.
In fact by making a change of variables
\ba
x_0=x_1x_2, \hskip3mm \zeta=\left({x_1\over x_2}\right)^{12}
\ea
(\ref{defining}) is rewritten as
\ba
&&0={B'\over 12}x_0^{12}
+\frac1{12}x_3^{12}+
\frac13x_4^3+\frac12 x_5^2
-\psi_0 x_0x_3x_4x_5
-\frac16\psi_1(x_0x_3)^6\no \\
&&-\frac14\psi_3(x_0x_3x_4)^2-\frac14 \psi_4 (x_0x_3)^4 x_4-\frac13\psi_5(x_0x_3)^3x_5.
\ea
where 
\ba
B'(\zeta)={1\over 2}\left(B\zeta+{B\over \zeta}-2\right).
\ea
We see the structure of $K3$ surfaces (in $WCP_{12}(1,1,4,6)$)
fibered  over $\P^1$ parametrized by $\zeta$.

Change of variables $\{x_0,x_3,x_4, x_5\}$ induce transformations among the parameters $\{\psi_i\}$ and we  may choose a gauge where 
$\psi_0=0,\psi_3=0,\psi_5=0$.
Then $B$, $\psi_1$, $\psi_4$ parametrize the complex structure moduli of the manifold.

If one further introduces a change of variables \ba
\xi &=\left(\displaystyle{\frac{x_3}{x_0}}\right)^6,\hskip3mm
x=\displaystyle{\frac{x_4}{(x_0x_3)^2}},\hskip3mm
y=\frac{x_5}{(x_0x_3)^3},
\ea
one can see the elliptic fibration over  the base 
parametrized by $\zeta$ and $x$ \begin{equation}
0=\frac12 y^2+\frac1{12} \left(\xi + \frac{B'(\zeta))}\xi \right)+\frac16P(x)\label{fibereq}
\end{equation}where\begin{align}
P(x)&=2x^3-\frac32\psi_4 x-\psi_1.
\end{align}

The elliptic curve degenerates when i)  $P^2-B'=0$ or ii) $B'=0$. 
These equations determine curves $\Sigma$, $\Sigma'$ inside the base, respectively.

\subsection{Geometric-Engineering Limit}
If we take the limit $B=2\epsilon \to 0$,
 then the K3 fiber acquires ADE singularities
everywhere on the base $\P^1$.
In the description we reviewed above,
 the curve $\Sigma'$ moves away to  $\zeta\to\pm\infty$ and 
we obtain the situation treated in the geometric engineering limit,
in which $SU(3)$ or $SU(2)\times SU(2)$
gauge theory decouples from gravity.
In order to see the field theory dynamics,
we need to take a controlled limit
so that when we parametrize the moduli using $a,b$ as\begin{equation}
\psi_4= 4 \epsilon^{2/3} a\quad\text{and}\quad
\psi_1=1+4\epsilon (b+{1\over 4}).
\end{equation} where $a$ and $b$ are kept finite.

In the limit we find the curve $\Sigma_+$, a branch of $\Sigma$, given by  \begin{equation}
\Sigma_+:\quad P(x)+ \sqrt {B'(\zeta)}=0
\end{equation} is reduced in the lowest order in $\epsilon$ to
\begin{equation}
\Sigma_+: \quad 0=\frac 14(\zeta+\frac1\zeta)+\tilde x^3-3a\tilde x-2 b-{1\over 2}
\label{sigmaminus}
\end{equation} where $x=\epsilon^{1/3}\tilde x$.
This is precisely the Seiberg-Witten curve for the pure $SU(3)$ gauge theory.
One can check that the integral of the holomorphic three-form
on the three-cycle corresponding to the gauge theory
is reduced to that of the
Seiberg-Witten differential on one-cycle on $\Sigma_+$.

\def\va{{v_a}}\def\vb{{v_b}}
\def\ta{{t_a}}\def\tb{{t_b}}
Let us concentrate in the following the patch in the moduli space where
$\epsilon$ is small and $a$ and $b$ is finite.
We take the basis of the 3-cycles as in eq. (6.31) in the section 6.4 in ref \cite{Billo},
\begin{equation}
(V_\va,V_\vb,V_\ta,V_\tb,T_\va,T_\vb,T_\ta,T_\tb)\label{basis}.
\end{equation}
$V_\va$, $V_\vb$, $T_\va$ and $T_\vb$ give the periods of the $SU(3)$ gauge theory,
while the rest are the ones we need to embed the gauge theory into supergravity.
We call the former the field theory periods, and the latter the supergravity periods.
Their intersection form is given in eq. (6.32) of ref \cite{Billo} and is reproduced
in the appendix \ref{explicitmonodromies}.

The singular loci in this patch are i) $\epsilon=0$ and ii) the singular locus for the pure $SU(3)$ theory.
The monodromy of the cycles around the locus $\epsilon=0$
is given by the formula (7.40) of \cite{Billo}
and the complex structure $\Omega$ transforms
as $\Omega\rightarrow  \omega \Omega$ where 
$\omega=e^{2\pi i/3}$. Making use of these information one can work out the
Jordan decomposition and we see that the periods behave as (we denote the period by the same symbol as the corresponding cycle)
\begin{align}
V_\va, V_\vb, T_\va, T_\vb &\sim \epsilon ^{1/3}\\
V_\ta, V_\tb, T_\ta,T_\tb &\sim \text{const} + \text{const}'\,\frac{-2}{2\pi i} \log \epsilon ,
\end{align}i.e. the behavior of the periods is of  Case II.
Thus, the integral of the Ashok-Douglas density converges
around this limit.

We see that the monodromy,
which is unipotent only after being cubed, makes
four of the periods parametrically small. 
Let us recall that the mass squared of a BPS saturated particle with
central charge $\Omega_I$ is given by
\begin{equation}
m_I^2=e^K |\Omega_I|^2.
\end{equation}Then, the mass scale $M$
of the solitons corresponding to the supergravity periods
is given by \begin{equation}
M\sim \text{const.} 
\end{equation}
 The ratio
between the dynamical scale $\Lambda$ of the gauge theory and the 
mass scale $M$ of ambient supergravity is given by, \begin{equation}
\frac\Lambda{M} \sim \frac{\epsilon^{1/3}}{\log\epsilon}
\end{equation}
If one first approximates the degeneration
by the ALE fibration over the sphere, as usually done in
the geometric engineering, one cannot capture
the logarithmic dependence on $\epsilon$.
As we saw in the previous sections, logarithmic behaviors
in the periods is the key determining the 
properties of the index density.
Thus in order to study the distribution of the
vacuum, one has to start from the compact Calabi-Yau manifolds
before taking their degenerate  non-compact limit.
The appearance of the logarithm is expected since
the component $\epsilon=0$ of the boundary divisor
intersects with the large complex structure limit
point, where the monodromy is maximally unipotent.

\subsection{Argyres-Douglas point}
Next we consider what will happen near the Argyres-Douglas (AD) point
in the $SU(3)$ moduli space, now embedded in the
Calabi-Yau moduli space at some small $\epsilon\ll 1$.
We know how the monodromy acts on the field theory periods.
However, it is necessary to determine its action also on the 
supergravity periods.

The Argyres-Douglas point is where two mutually non-local cycles
simultaneously vanish.
When $a$ and $b$ are both small,
the curve $\Sigma_+$ \eqref{sigmaminus}  becomes
\begin{align}
\Upsilon:&\quad w^2=\tilde x^3-3a\tilde x- 2b
\end{align} where $\zeta=1+2 i w$.
In the following we focus on the two dimensional
moduli space parametrized by $a$ and $b$ near the origin.

\paragraph{Structure of the singular loci}
Let us first consider the structure of the singular locus.
The elliptic curve $\Upsilon$ degenerates when $a^3=b^2$.
Away from the origin $(a,b)=(0,0)$,  this is just a conifold locus
entangled inside $\C^2$ spanned by $(a,b)$.
The origin corresponds to the Argyres-Douglas point.

The origin is not normal crossing in this coordinate system.
We need to blow up the moduli space at the origin three times
consecutively in order to obtain a normal-crossing boundary divisor (for details of the blowup procedure see appendix B).
The final coordinate system $(s,\alpha)$ is related to the original one by
the transformation\begin{equation}
a=s^2\alpha, \qquad b=s^3\alpha.
\end{equation} 

It is known that the Seiberg-Witten differential
in this limit is proportional to $wd\tilde x$.  This in turn means
that $s$ is precisely the scaling variable for the conformal theory
at the Argyres-Douglas point.
We can also see that, by scaling with $w=s^{3/2}\hat w$ and $\tilde x=s\hat x$,
$\alpha$ controls the shape of the elliptic curve $\Upsilon$.
Thus we see that the coordinate system where
the boundary divisor is normal crossing is the usual one
used by field theorists to capture the physics of the system.

Three exceptional cycles are introduced during the process of blowups.
These cycles, combined with the lift of
the original conifold locus, constitute the boundary divisor.
It is convenient to introduce another variable $t=s\alpha$.
Then the singular loci are i)  $D_2:\ t=0$ which is $\P^1$,
ii) $D_3:\ s=0$ which is also $\P^1$,
iii) another $\P^1$ parametrized by
$\alpha=t/s$ which we call $D_{AD}$, and finally
iv) $D_c:\  s=t$, the lift of the original $a^3=b^2$ locus
which  is connected to the other parts of the moduli space.
The intersection among them is depicted in figure \ref{blowupofcusp} in appendix \ref{appB}.

On $D_2$, the complex structure $\tau$ of the $\Upsilon$ is $i$,
thus its automorphism group has order two. On $D_3$, $\tau$ is
$e^{2\pi i/6}$ and the automorphism group is $\Z_3$.
These automorphism is reflected on the moduli space
as the orbifold singularity of corresponding order.
Finally on $D_c$ one of the cycle of the torus degenerates.

\paragraph{Monodromies}

Two one-cycles on $\Upsilon$ corresponds to the three-cycles
$A=V_\va$ and $B=V_\vb$.  Actions of monodromy on $A$ and $B$
around various boundary components are given by:
\begin{align}
M_{AD}(A,B)^T&= (-A,-B) ^T& &\text{around}\ D_{AD},\\
M_{2}(A,B)^T&= (-B,A)^T & &\text{around}\ D_{2},\\
M_{3}(A,B)^T&= (A+B,-A)^T && \text{around}\ D_{3},\\
M_{c}(A,B)^T&= (A+B,B) ^T& &\text{around}\ D_{c}.
\end{align}
The first three of monodromies are idempotent, $M_{AD}^2=M_2^4=M_3^6=id$.
Only the last monodromy contains the logarithmic one.
In this description, two mutually non-local
solitons becoming massless at the AD point is attributed
to the monodromy around $D_{AD}$ which flips the sign of both 
$A$ and $B$. This means $A$ and $B$ must be zero
on $D_{AD}$.

We need to check the action of monodromies on the other periods.
This is rather cumbersome, as the four 3-cycles
$T_\va$, $T_\vb$, $T_\ta$ and $T_\tb$ are pinched 
at $a=b=0$.  One can draw the picture of the action of the monodromy
on the cuts of the differentials and then determine the monodromy on
the cycles, but this is rather tedious.
A better way is to first check that, by drawing pictures,
the monodromy of an arbitrary cycle
$C$ is given by the formula\begin{equation}
C\to C+ p_C A+ q_C B, \label{genPL}
\end{equation} 
that is, any cycle is shifted by the linear combination of the vanishing
cycles.  Then one can fix the coefficients $p_C$ and $q_C$
by demanding that the monodromy conserves the intersection number of
two cycles, given the monodromy action on $A$ and $B$.  
(\ref{genPL}) appears to be a generalization of Picard-Lefschetz formula 
for the case of simultaneously vanishing cycles. 

Hence we have a homomorphism from $\Sp(2,\Z)$ acting on $A$, $B$
to $\Sp(8,\Z)$ acting on all the periods of the Calabi-Yau manifold.
For the completeness,
we presented the monodromy matrices in the appendix \ref{explicitmonodromies}.
They satisfy $[M_2,M_{AD}]=[M_3,M_{AD}]=[M_c,M_{AD}]=0$, as they should.
Since $M_2^4=M_3^6=M_{AD}^2=id.$,  after taking four-, six- or two-folded cover 
of the original punctured disk $\Delta^*$ there remains no non-analytic
behavior of periods around $D_{AD}$, $D_2$, and $D_3$.
Then the only possible divergence comes from the conifold locus $D_c$.
We have seen, however, that they do not cause the divergence of the index density.
Thus the situation of the AD point is the same as the conifold case.

\section{Conclusions}

We have studied the behavior of distribution of flux vacua around singular loci in CY complex structure moduli space in type IIB string theory.
We have shown in  various cases of singularities the distribution is normalizable, 
which roughly means that
there exist only a finite number of vacua around each singular locus. This observation may be of some use
in the future discussion of vacuum selection in superstring theories.

Although we have discussed individual cases separately in this article, we feel that there should be a more unified and rigorous treatment to show the normalizable behavior of density distributions.
It seems that the key is to bring the singularity to a normal crossing form.
Fortunately, it is known that this is always possible
by fundamental mathematical theorems
on the resolution of singularities.

\paragraph{Acknowledgment}
Research of T.E. is supported in part by Japan Ministry of Education, Culture and Sports under contracts no.15540253,16081206.
Research of Y.T. is supported by the JSPS predoctoral fellowship.
Both are grateful to the hospitality at KITP during the workshop ``Mathematical Structures in String Theory''
  where this work has been completed.
The work is partially supported
 by the National Science Foundation under Grant No. PHY99-07949.

\appendix

\section{Facts on the variation of the Hodge structures}
\label{HodgeStructure}
We present in this appendix standard definitions and facts
on the variation of the Hodge structures\cite{Schmid}
which we have used in the main part of this paper.
For brevity, we restrict to the case of Calabi-Yau three-folds.
We abbreviate $h_{1,2}=n$.

\subsection{Hodge Structures}
Let $L=\Z^{2n+2}$ be a lattice with a skew symmetric bilinear form $\eta$.
A Hodge structure\footnote{of weight three.
The weight three reflects the fact that we consider a three-fold.
We omit the weight in the following unless necessary.}
on $L$ is a flag \begin{equation}
L\otimes \C=F^0\supset F^1 \supset F^2\supset F^3\supset 0
\end{equation}with dimensions \begin{equation}
\dim F^1=2n+1,\quad \dim F^2=n+1,\quad \dim F^3=1 \label{flag}
\end{equation} such that
\begin{equation}
\bar F^1\cup F^3 = \bar F^2 \cup F^2= L \otimes \C.\label{realityofHS}
\end{equation} For such a flag, we define as
$H^{p,q}=F^p\cap \bar F^q$.
The Weil operator $C$ for a given Hodge structure
is defined to be the multiplication by
$i^{p-q}$ on elements of $H^{p,q}$.
A Hodge structure is called polarized with respect to $\eta$ if
\begin{equation}
\eta(H^{p,q}, H^{r,s})\ne 0 \qquad \text{only if}\quad p=s,\ q=r,
\end{equation} and \begin{equation}
\eta(Cv,\bar w) \ \text{is a positive definite Hermitean form.}
\end{equation}
A Hodge structure is sometimes called a pure Hodge structure
in order to distinguish from a mixed Hodge structure,
which will be introduced later.

For a fixed $L$ and $\eta$, we denote by $D$
the set of all Hodge structures on it.
$\Sp(2n+2,\R)$ acts on $D$  transitively and thus
$D$ can be expressed as \begin{equation}
D\simeq \Sp(2n+2,\R)/V
\end{equation}where $V$ is the  compact subgroup which fixes a flag chosen 
as the basepoint. $D$ is known as the Griffiths' period domain.
We will later
make use of the set $\check D$ of all flags \eqref{flag} which might not satisfy
\eqref{realityofHS} .  $\check D$  can be expressed as \begin{equation}
\check D\simeq \Sp(2n+2,\C)/B
\end{equation} where $B$ is the subgroup which fixes the basepoint flag.
$D$ and $\check D$ carries the tautological flag
bundle $F^0\supset F^1 \supset F^2\supset F^3$. 
They are the so-called Hodge bundles.
It is known that $D$ is a K\"ahler manifold.

We call a holomorphic map $f$ from a K\"ahler manifold $M$
to $D$ horizontal
if the covariant derivative of any section of the pullback of $F^p$
is in $F^{p-1}$.  This condition can be translated so that
$df$ maps $TM$ inside the horizontal subbundle $H\subset TD$. 
$H$ is also a Hermitean  bundle.

All these properties and definitions are abstracted from the
variation of $F^p=\oplus_{p\le q} H^{q,3-q}$ inside $H^3(X,\Z)\otimes \C$.
Thus the period map of the Calabi-Yau moduli space
determines the horizontal mapping into the Griffiths' period domain $D$.

\subsection{Variation of Hodge Structures}\label{VHS}
Let us consider a polarized Hodge structure 
on a half plane P defined by Re$t>0$
with a horizontal map $f:P\rightarrow D$.
Furthermore suppose $f(t+2\pi i)=\gamma f(t)$ for an 
integral matrix
$\gamma\in \Sp(2n+2,\Z)$.
Then, it reduces to a holomophic map $f\circ \pi: \Delta^* \to \Gamma\backslash D$
by composing $\pi: t\mapsto z=e^{-t}$. This is called a
variation of the Hodge structure on $\Delta^*$ with monodromy $\gamma$. 
It is known that $\gamma$ is quasi-unipotent, that is,
there are some integers $s$ and $r$ such that
$(\gamma^s-1)^r=0$. A sketch of the proof is provided in the
next subsection, \ref{monodromytheorem}.

By redefining $\gamma^s$ as $\gamma$ , one can assume that
the monodromy $\gamma$ is unipotent.
Let us denote the logarithm of $\gamma$ by $N$.  $N=\log\gamma$ is nilpotent.
Then, the map $\exp(-tN/2\pi i) f (t)$ from  $P$ to $\check D$ becomes
periodic and determines the single-valued map $g(z)$
from $\Delta^*$ to $\check D$.
An important subtlety here is that after the multiplication
by $\exp(-tN/2\pi i)$
the flags no longer  satisfy the conditions \eqref{realityofHS}.
Thus, the map is not to $D$ but to $\check D$. 
The basic theorem is
\paragraph{Nilpotent orbit theorem of Schmid:}
the map $g(z)$ can be holomorphically extended to
the disk $\Delta\supset \Delta^*$ including the origin $z=0$.

\bigskip

In particular, on a coordinate patch near $g(0)$, we see that
the periods $\Omega_I$ have the expansion of the form
\begin{equation}
\Omega_I=\exp(\frac{N}{2\pi i}\log z )
(\Omega_I^{(0)}+\Omega_I^{(1)} z+ \Omega_I^{(2)}z^2+\cdots),
\label{expansionofOmega}
\end{equation} where $\Omega_I^{(0)}$ is nonzero.

Let us denote the filtration corresponding to $g(0)\in \check D$ 
by $F^0_\infty\supset F^1_\infty\supset F^2_\infty\supset F^3_\infty$.
This is not a Hodge structure in general.
However, it still holds that \begin{equation}
NF^p_\infty\subset F^{p-1}_\infty.\label{morphism}
\end{equation}
This is basically
because $e^{N}$ can be obtained by integrating the horizontal
connection $\nabla F^p\subset F^{p-1}\otimes \Omega^{(1,0)}\M$ around $z=0$.
From \eqref{morphism} we conclude $N^4=0$.

Let us introduce another filtration\begin{equation}
0\subset W_{0}\subset W_{1}\subset\cdots\subset W_{5}\subset
W_{6}=L\otimes\C
\end{equation} using the nilpotent part $N$ such that 
$L\otimes \C$ is a representation of $SL(2)$ with $N$ representing
the lowering operator $J^-$ and with $W_j$ the span of vectors
with $J_z$ eigenvalues equal or less than $(j-3)/2$.
Note that the subscript $j$ is restricted in $0\le j\le 6$ because
$N^4=0$.
We call the kernel of $N^{j+1}$ on $W_{3+j}$ the primitive part of the filtration
$W_*$ and denote it by $P_{3+j}$.

The pair of the filtrations $F_\infty^p$ and 
$W_l$ constructed above satisfies the following
fundamental properties:
\begin{multline}
\text{For each $l$, the filtration}\ F^p _\infty \cap W_l/F_\infty^p\cap W_{l-1}
\ \text{on $W_l/W_{l-1}$}\\
\text{ is a pure Hodge structure of weight $l$,}\label{MHS}
\end{multline}and furthermore,\begin{multline}
\text{For each $l\ge 0$, the filtration}\ F^p_\infty\cap P_{l+3}\ \text{on $P_{l+3}$ }\\ 
 \text{is a pure Hodge structure of weight $l+3$ polarized with respect to
$\eta( \cdot,N^{l}\cdot)$.}
\label{pMHS}
\end{multline}
A pair of filtrations $F_\infty^p$ and $W_l$ of $L\otimes \C$
satisfying the condition \eqref{MHS} is called a mixed Hodge structure.
If it also satisfies the condition \eqref{pMHS}, it is said to be
polarized with respect to the bilinear form $\eta$ on $L$.
The mixed Hodge structure constructed from the
variation of the Hodge structure in the way
just described is called the limiting
mixed Hodge structure.
The fact that the limiting mixed Hodge structure is polarized
gives a strong control on the growth of the norm of the
periods.

Let us estimate the growth of $e^{-K}=i\eta \bar\Omega\Omega$
near the singular locus using the property of the mixed Hodge structure.
The expansion \eqref{expansionofOmega} tells us that
\begin{align}
e^{-K}&\sim i\eta(\exp(- t\frac{N}{2\pi i})
\Omega^{(0)})^* \exp(-t\frac{N}{2\pi i})\Omega^{(0)}\nonumber\\
&=i\eta \overline{\Omega^{(0)}}\exp(- (\Re t) \frac{N}{\pi i})\Omega^{(0)}.
\end{align}
Recall $NF^p_\infty \subset F^{p-1}_\infty$ and 
$\Omega^{(0)}$ is in $F^3_\infty$ by definition.
Thus, $\Omega^{(0)}$ is primitive under the action of $N$.
Hence, $\Omega^{(0)}\in P^{q}$ where $q$ is an integer
such that $N^q\Omega^{(0)}\ne 0$ but $N^{q+1}\Omega^{(0)}=0$.

Thus,\begin{equation}
i\eta \overline{\Omega^{(0)}}\exp( -(\Re t) \frac{N}{\pi i})\Omega^{(0)}
\sim c (\Re t)^q
\end{equation}  where the proportionality constant \begin{equation}
c= i\eta \overline{\Omega^{(0)}}\left(\frac{N}{2\pi i}\right)^q\Omega^{(0)}
\end{equation} is
guaranteed to be nonzero from the condition \eqref{pMHS}.

\subsection{Sketch of the Proof of the Monodromy Theorem}
\label{monodromytheorem}
We reproduce here the proof of the monodromy theorem. The proof  is originally due to A. Borel.

We use a kind of generalized Schwarz' theorem
which governs the behavior of holomorphic maps
between spaces with bounded curvatures. One useful version is
\cite{YauSchwarz}

\paragraph{Theorem (Yau)}
Let $M$ and $N$ be Hermitean manifolds, with the Ricci curvature of $M$ 
bounded from below,
and the holomorphic bisectional curvature of $N$ bounded from above
by a negative number, where the holomorphic bisectional curvature is defined as
$R_{i\bar j k\bar l} u^i \bar u^{\bar j} v^k \bar v^{\bar l}$ for
two vectors $u$ and $v$ with unit length.
Then there is a number $K$ depending 
only on the two bounds such that 
we have $f^*(ds^2_N) < K ds^2_M$ for any holomorphic mapping $f:M\to N$.

\bigskip

Assuming this, the proof of the monodromy theorem
is not so difficult.
Embed a punctured disk $\Delta^*$
parametrized by $z$ inside the Calabi-Yau moduli space so that
it describes a one-parameter deformation around a singular locus,
and let us denote by $P$ the half plane $t>0$,
which is the universal cover of $\Delta^*$ by the relation $z=e^{-t}$.
We endow $P$ with the standard Poincar\'e metric.
The period map determines a horizontal mapping from $P$ to $D$.
Let us call it $f$.

We first need to show by direct calculation that the Ricci curvature of 
the horizontal subbundle satisfies $\Ric_H < -\omega_D$. 
Let us denote by $W$ the image of the upper half plane $P$ under $f$.
Since $TW$ is a holomorphic subbundle of $H$,
we have $\Ric_W\le \Ric_H$.  Then
we  let $M=P$ and $N=W$ and apply the theorem above.
Thus we  find that the map $f$ decreases the distance 
by a constant multiple.

Consider two sequences of points $t=m$ and $m+2\pi i$ in $P$ ($m=1,2,3,\ldots$)
and denote their images under $f$
by $g_mV$ and $g'_m V\in D\simeq G/V$,
respectively.  Recall the shift $t\to t+2\pi i$ in $P$ generates the monodromy
around the puncture in $\Delta^*$. Thus, $g'_m=\gamma g_m$ when
we denote the monodromy matrix by $\gamma$.
If we combine the above theorem and the fact
that the distance between $t=m$ and $t=m+2\pi i$ is $1/m$, we find that
the distance between $g_mV$ and $g'_mV$ is less than $\mbox{const}/m$.
Thus $g_m^{-1}\gamma g_m$ asymptotes to the compact subgroup $V$.
Thus all the absolute values of the eigenvalues of $\gamma$  must be one.
Finally recall that $\gamma$ is a matrix with integer  entries,
which in turn means that the eigenvalues are algebraic integers.
From the Kronecker's theorem, which says that 
an algebraic integer is a root of unity if
all the absolute values  of its conjugates are one, we conclude the
eigenvalues of $\gamma$ to be some roots of unity.

The proof of the theorem of Kronecker is also easy. 
Consider an algebraic integer $\alpha$
whose conjugates all have absolute value one.  Suppose that it satisfies
an degree $n$ monic polynomial equation. Then, the absolute values
of the coefficients of the polynomial, which should be integers,
 are also bounded.  Thus, total number
of such algebraic integer is finite.  Since $\alpha^k$ for any $k$ also satisfies 
some degree $n$ monic polynomial equation and the absolute values
of all its conjugates are one, we need to have $\alpha^k=\alpha^{k'}$ for some
$k\ne k'$. Thus $\alpha$ is some root of unity.

\section{Blowup of the Cusp $a^3=b^2$ in Detail}
\label{appB}
First recall that blowing up the origin of the plane $\C^2$
with coordinates $(x,y)$ replaces the origin by the projective plane
$\CP^1$ which describes in which angles one is approaching the origin.
Denoting the homogenious coordinates of the $\CP^1$ with $[\xi:\eta]$,
the total space of the blowup is \begin{equation}
\{  ([\xi:\eta],(x,y))\in \CP^1\times \C^2\  |\ \xi y = \eta x\}.
\end{equation}
Near $[1:0]\in\CP^1$ we can use $x$ and $k=\eta/\xi$ as the local
coordinates of the blowup. Hence in this patch blowup essentially means
the coordinate change from $(x,y)$ to $(x,y/x)$.
$\CP^1$ at $x=y=0$ is called the exceptional curve.

Consider a curve $C$ in $\C^2$ passing through the origin.
The inverse image of the curve contains the exceptional curve.
The proper transform of the curve is defined as the
closure of the inverse image of $C\setminus\{0\}$,
and this can be determined by doing the coordinate change
from $(x,y)$ to $(x,k)$ in the defining equation of the curve
and throwing away the component describing the exceptional curves.

After recalling these fundamentals, let us blow up the cusp $a^3=b^2$
in detail.
\paragraph{1st blowup.}
Introduce a $\P^1$ at the origin $(a,b)=(0,0)$ with
coordinates $[1:s]$ and call it $E_1$.   Now $b=sa$. The defining equation becomes
$a^2(a - s^2)=0$. $a=0$ defines an exceptional curve $E_1$.
The proper transform of the cusp is $a=s^2$.
\paragraph{2nd blowup.}
Introduce another $\P^1$ at the origin $(a,s)=(0,0)$ with
coordinates $[t:1]$ and call it $E_2$.
Now $a=st$. The equation $a=s^2$ becomes
$s(s-t)=0$. $s=0$ describes the exceptional curve $E_2$.
The proper transform of the curve $a=s^2$ is $s=t$.
The proper transform of $E_1$ can be calculated in the same way,
and we obtain $t=0$.
\paragraph{3rd blowup.}
In order to split the intersection of curves $s=t$, $s=0$ and $t=0$ at the origin $(s,t)=(0,0)$ we 
introduce a parameter $\alpha$ defined by $t=s\alpha$.
$\alpha$ parametrizes a still another $\P^1$ called $E_3$. 
Then the proper transform of the curve $s=t$ intersects $E_3$ transversally at $\alpha=1$ and 
those of $E_{1,2}$ intersect $E_3$ transversally at $\alpha=0$
and $\alpha=\infty$, respectively. 

\begin{figure}
\includegraphics[width=.9\textwidth]{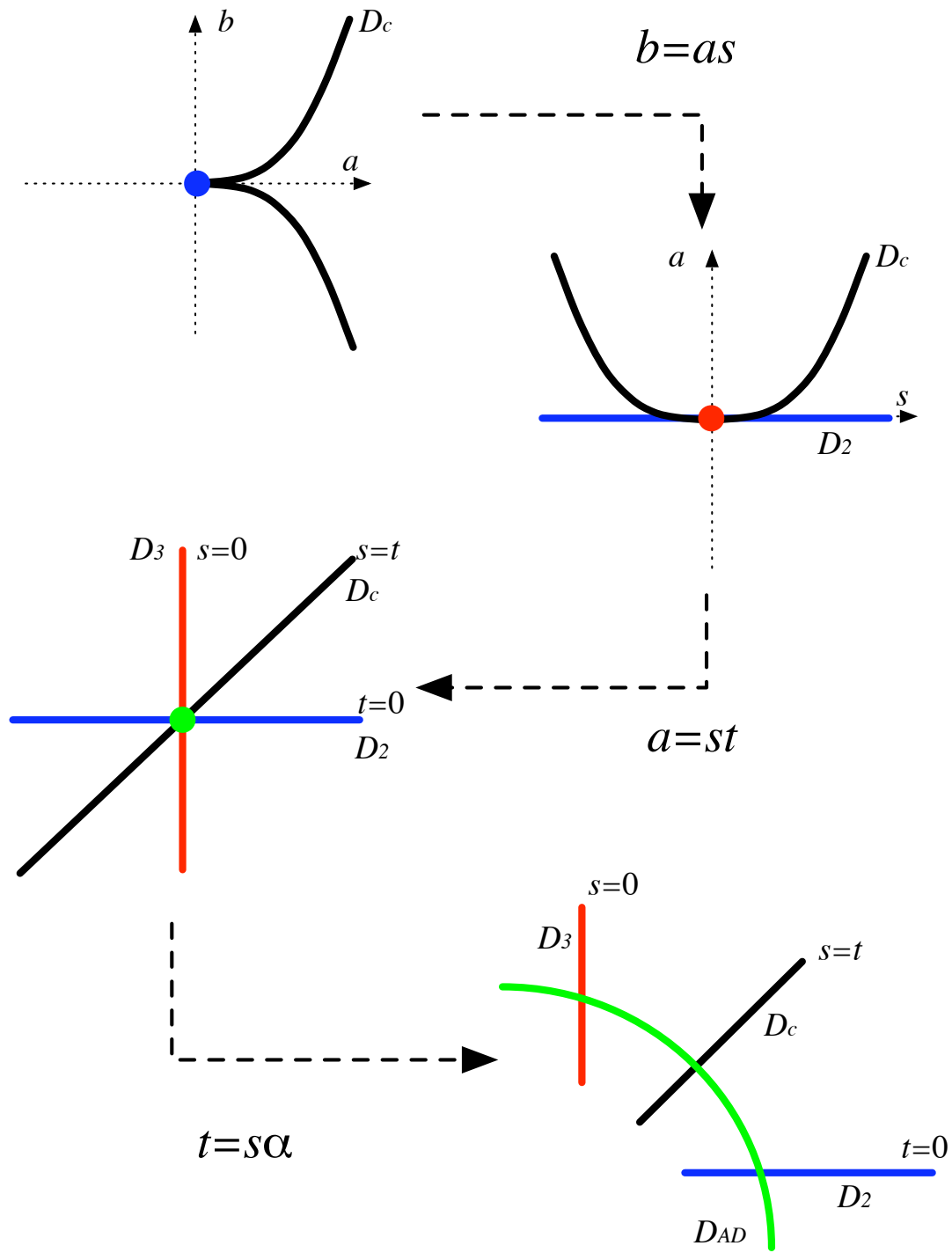}
\caption{Blowing up of the cusp $b^2=a^3$\label{blowupofcusp}}
\end{figure}

Combining all these steps
and renaming as $D_2=E_1$, $D_3=E_2$ and $D_{AD}=E_3$,
we arrive at the figure \ref{blowupofcusp}.

\section{Explicit forms of the monodromy matrices}\label{explicitmonodromies}
The intersection form of the cycles
is
\begin{equation}
\begin{pmatrix}
 0&-1&0&0&-2&1&0&-1 \\
  1&0&0&0&1&-2&1&0 \\
 0&0&0&-2&0&1&0&0 \\ 
 0&0&2&0&-1&0&0&0 \\
 2&-1&0&1&0&0&0&0 \\
  -1&2&-1&0&0&0&0&0 \\ 
 0&-1&0&0&0&0&0&0\\
  1&0&0&0&0&0&0&0
\end{pmatrix},
\end{equation}which is given in eq. (6.32) of ref. \cite{Billo}.

Let us denote the monodromy matrix around the locus $D_*$ 
by $M_*$. 
It is understood they act as \begin{equation}
(V_\va,\ldots,T_\tb)^T\to 
M(V_\va,\ldots,T_\tb)^T.
\end{equation} where the order and choice of the basis is as in \eqref{basis}.
Here are the $8\times8$ matrices:
\begin{align}M_3&=
\begin{pmatrix}
1&1&0&0&0&0&0&0\\
-1&0&0&0&0&0&0&0\\
0&0&1&0&0&0&0&0\\
0&0&0&1&0&0&0&0\\
-2&-1&0&0&1&0&0&0\\
1&-1&0&0&0&1&0&0\\
0&1&0&0&0&0&1&0\\
-1&-1&0&0&0&0&0&1
\end{pmatrix},&
M_2&=\begin{pmatrix}
0&-1&0&0&0&0&0&0\\
1&0&0&0&0&0&0&0\\
0&0&1&0&0&0&0&0\\
0&0&0&1&0&0&0&0\\
1&-3&0&0&1&0&0&0\\
1&3&0&0&0&1&0&0\\
-1&-1&0&0&0&0&1&0\\
1&-1&0&0&0&0&0&1
\end{pmatrix},\\
\null\nonumber \\
M_c&=\begin{pmatrix}
1&1&0&0&0&0&0&0\\
0&1&0&0&0&0&0&0\\
0&0&1&0&0&0&0&0\\
0&0&0&1&0&0&0&0\\
0&1&0&0&1&0&0&0\\
0&-2&0&0&0&1&0&0\\
0&1&0&0&0&0&1&0\\
0&0&0&0&0&0&0&1
\end{pmatrix},&
M_{AD}&=\begin{pmatrix}
-1&0&0&0&0&0&0&0\\
0&-1&0&0&0&0&0&0\\
0&0&1&0&0&0&0&0\\
0&0&0&1&0&0&0&0\\
-2&-4&0&0&1&0&0&0\\
4&2&0&0&0&1&0&0\\
-2&0&0&0&0&0&1&0\\
0&-2&0&0&0&0&0&1
\end{pmatrix}.
\end{align}


\begin{thebibliography}{99}

    \bibitem{BussoPolchinski}

  R.~Bousso and J.~Polchinski,
  ``Quantization of four-form fluxes and dynamical neutralization of the
  cosmological constant,''
 {\slshape   JHEP }{\bf 0006} (2000) 006
  [arXiv:hep-th/0004134].
 

    \bibitem{GiddingsPolchinski}

  S.~B.~Giddings, S.~Kachru and J.~Polchinski,
  ``Hierarchies from fluxes in string compactifications,''
 {\slshape   Phys.\ Rev.\ D }{\bf 66} (2002) 106006
  [arXiv:hep-th/0105097].
 

    \bibitem{KKLT}

  S.~Kachru, R.~Kallosh, A.~Linde and S.~P.~Trivedi,
  ``De Sitter vacua in string theory,''
 {\slshape   Phys.\ Rev.\ D }{\bf 68} (2003) 046005
  [arXiv:hep-th/0301240].
 

    \bibitem{statistics}

  M.~R.~Douglas,
  ``The statistics of string / M theory vacua,''
 {\slshape   JHEP }{\bf 0305} (2003) 046
  [arXiv:hep-th/0303194].
 
    \bibitem{Sethi}
     K.~Dasgupta, G.~Rajesh and S.~Sethi,
  JHEP {\bf 9908} (1999) 023
  [arXiv:hep-th/9908088];
   S.~Sethi, talk at KITP, UCSB, Nov 20, 1999 [{\tt http://online.kitp.ucsb.edu/online/susy\_c99/sethi/}].

    \bibitem{BetterRacetrack}

  F.~Denef, M.~R.~Douglas and B.~Florea,
  ``Building a better racetrack,''
 {\slshape   JHEP }{\bf 0406} (2004) 034
  [arXiv:hep-th/0404257].
 

    \bibitem{BarrenLandscape}

  D.~Robbins and S.~Sethi,
  ``A barren landscape,''
 {\slshape   Phys.\ Rev.\ D }{\bf 71} (2005) 046008
  [arXiv:hep-th/0405011].
 

    \bibitem{TripathyTrivedi}

  P.~K.~Tripathy and S.~P.~Trivedi,
  ``D3 brane action and fermion zero modes in presence of background flux,''
 {\slshape   JHEP }{\bf 0506} (2005) 066
  [arXiv:hep-th/0503072].
 

    \bibitem{KalloshIndex}

  E.~Bergshoeff, R.~Kallosh, A.~K.~Kashani-Poor, D.~Sorokin and A.~Tomasiello,
  ``An index for the Dirac operator on D3 branes with background fluxes,''
  arXiv:hep-th/0507069.
 

    \bibitem{LustTripathy}

  D.~L\"ust, S.~Reffert, W.~Schulgin and P.~K.~Tripathy,
  ``Fermion zero modes in the presence of fluxes and a non-perturbative
  superpotential,''
  arXiv:hep-th/0509082.
 

    \bibitem{AshokDouglas}

  S.~Ashok and M.~R.~Douglas,
  ``Counting flux vacua,''
 {\slshape   JHEP }{\bf 0401} (2004) 060
  [arXiv:hep-th/0307049].
 

    \bibitem{DenefDouglas}

  F.~Denef and M.~R.~Douglas,
  ``Distributions of flux vacua,''
 {\slshape   JHEP }{\bf 0405} (2004) 072
  [arXiv:hep-th/0404116].
 

    \bibitem{Zelditch}

  M.~R.~Douglas, B.~Shiffman and S.~Zelditch,
  ``Critical points and supersymmetric vacua. III: String/M models,''
  arXiv:math-ph/0506015.
 

    \bibitem{BlackCondensation}

  B.~R.~Greene, D.~R.~Morrison and A.~Strominger,
  ``Black hole condensation and the unification of string vacua,''
 {\slshape   Nucl.\ Phys.\ B }{\bf 451} (1995) 109
  [arXiv:hep-th/9504145].
 

    \bibitem{GeometricEngineering}

  S.~Katz, A.~Klemm and C.~Vafa,
  ``Geometric engineering of quantum field theories,''
 {\slshape   Nucl.\ Phys.\ B }{\bf 497} (1997) 173
  [arXiv:hep-th/9609239].
 

    \bibitem{AD}

  P.~C.~Argyres and M.~R.~Douglas,
  ``New phenomena in SU(3) supersymmetric gauge theory,''
 {\slshape   Nucl.\ Phys.\ B }{\bf 448} (1995) 93
  [arXiv:hep-th/9505062].
 

    \bibitem{NewSCFT}

  P.~C.~Argyres, M.~R. Plesser, N.~Seiberg and E.~Witten,
  ``New N=2 Superconformal Field Theories in Four Dimensions,''
 {\slshape   Nucl.\ Phys.\ B }{\bf 461} (1996) 71
  [arXiv:hep-th/9511154].
 

    \bibitem{EHIY}

  T.~Eguchi, K.~Hori, K.~Ito and S.~K.~Yang,
  ``Study of $N=2$ Superconformal Field Theories in $4$ Dimensions,''
 {\slshape   Nucl.\ Phys.\ B }{\bf 471} (1996) 430
  [arXiv:hep-th/9603002].
 

    \bibitem{Taxonomy}

  A.~Giryavets, S.~Kachru and P.~K.~Tripathy,
  ``On the taxonomy of flux vacua,''
 {\slshape   JHEP }{\bf 0408} (2004) 002
  [arXiv:hep-th/0404243].
 

    \bibitem{ConlonQuevedo}

  J.~P.~Conlon and F.~Quevedo,
  ``On the explicit construction and statistics of Calabi-Yau flux vacua,''
 {\slshape   JHEP }{\bf 0410} (2004) 039
  [arXiv:hep-th/0409215].
 

    \bibitem{DouglasLu}

  M.~R.~Douglas and Z.~Lu,
  ``Finiteness of volume of moduli spaces,''
  arXiv:hep-th/0509224.
 

    \bibitem{LuNatsukawa}


 Z.~Lu and E.~Natsukawa, ``On the Weil-Petersson Geometry of Calabi-Yau Moduli,'' arXiv:math.DG/0509172

    \bibitem{PlanckQuantization}

  E.~Witten and J.~Bagger,
  ``Quantization Of Newton's Constant In Certain Supergravity Theories,''
 {\slshape   Phys.\ Lett.\ B }{\bf 115} (1982) 202.
 

    \bibitem{StromingerSpecial}

  A.~Strominger,
  ``Special Geometry,''
 {\slshape   Commun.\ Math.\ Phys.\  }{\bf 133} (1990) 163.
 

    \bibitem{CernSpecial}

  L.~Castellani, R.~D'Auria and S.~Ferrara,
  ``Special K\"ahler Geometry: An Intrinsic Formulation From N=2 Space-Time
  Supersymmetry,''
 {\slshape   Phys.\ Lett.\ B }{\bf 241} (1990) 57.
 

    \bibitem{GVW}

  S.~Gukov, C.~Vafa and E.~Witten,
  ``CFT's from Calabi-Yau four-folds,''
 {\slshape   Nucl.\ Phys.\ B }{\bf 584} (2000) 69
  [Erratum-ibid.\ B {\bf 608} (2001) 477]
  [arXiv:hep-th/9906070].
 

    \bibitem{TaylorVafa}

  T.~R.~Taylor and C.~Vafa,
  ``RR flux on Calabi-Yau and partial supersymmetry breaking,''
 {\slshape   Phys.\ Lett.\ B }{\bf 474} (2000) 130
  [arXiv:hep-th/9912152].
 

    \bibitem{Todorov}

  A.~Todorov,
  ``Weil-Petersson volumes of the moduli spaces of CY manifolds,''
  arXiv:hep-th/0408033.
 

    \bibitem{LuSun1}


 Z.~Lu and X.~Sun, ``Weil-Petersson geometry on moduli space of polarized Calabi-Yau Manifolds,'' \textit{Journal de l'Institut Mathematique de Jussieu } \textbf{3} (2004) 185 [math.DG/0510020]

    \bibitem{LuSun2}


 Z.~Lu and X.~Sun, ``On the Weil-Petersson volume of the moduli space of Calabi-Yau manifolds,'' arXiv:math.DG/0510021

    \bibitem{Hayakawa}


 Y.~Hayakawa, ``Degenaration of Calabi-Yau Manifold with W-P Metric,'' arXiv:alg-geom/9507016

    \bibitem{Wang}


 C.-L.~Wang, ``On the incompleteness of the Weil-Petersson Metric Along Degenerations of Calabi-Yau Manifolds,'' \textit{Math. Research Lett. } \textbf{4} (1997) 157

    \bibitem{Viehweg}


 E.~Viehweg, ``Quasi-Projective Moduli for Polarized Varieties,'' \textit{Ergebnisse der Mathematik und ihrer Grenzgebiete, } 3. Folge, Band 30, Springer, 1995

    \bibitem{Billo}

  M.~Bill\'o, F.~Denef, P.~Fr\`e, I.~Pesando, W.~Troost, A.~Van Proeyen and D.~Zanon,
  ``The rigid limit in special K\"ahler geometry: From K3-fibrations to  special
  Riemann surfaces: A detailed case study,''
 {\slshape   Class.\ Quant.\ Grav.\  }{\bf 15} (1998) 2083
  [arXiv:hep-th/9803228].
 

    \bibitem{Schmid}


 W. Schmid, ``Variation of Hodge Structure: The Singularities of the Period mapping,'' \textit{Invent. Math. } \textbf{22} (1973) 211

    \bibitem{YauSchwarz}


 S.-T. Yau, ``A Generalized Schwarz Lemma for K\"ahler Manifolds,'' \textit{Amer. Jour. Math. } \textbf{100} (1978) 179

\end{thebibliography}
\end{document}